\documentclass[12pt]{article}
\usepackage[utf8]{inputenc}
\usepackage{textcomp} %non-italicised µ
\usepackage{graphicx}
\usepackage{wrapfig}
\usepackage{ulem}
\usepackage[dvipsnames]{xcolor}

\usepackage{amssymb}
\usepackage{amsmath}
\usepackage[a4paper, portrait, margin=1in]{geometry}
\usepackage{caption}
\usepackage{wrapfig}
\usepackage{color}

\title{All-optical control of phase singularities using strong light-matter coupling}
\author{Philip A. Thomas\footnote{p.thomas2@exeter.ac.uk}, Kishan S. Menghrajani and William L. Barnes\footnote{w.l.barnes@exeter.ac.uk}\\ \\ \small{Department of Physics and Astronomy, University of Exeter,} \\ \small{Exeter, EX4 4QL, United Kingdom}}
\date{}

\begin{document}

\maketitle

%\newpage

%%Nat Photon letters ask for a referenced 150-word introductory summary paragraph, not a separate abstract
\noindent \textbf{Strong light-matter coupling occurs when the coupling strength between a confined electromagnetic mode and a molecular resonance exceeds losses to the environment\cite{lidzey1998strong, torma2014strong, garcia2021manipulating}.
The study of strong coupling has been motivated by applications such as lasing\cite{ramezani2017plasmon} and the modification of chemical processes\cite{feist2017polaritonic}.
Here we show that strong coupling can be used to create phase singularities.
Many nanophotonic structures have been designed to generate phase singularities\cite{hoenig1991direct, kravets2008extremely, malassis2014topological, berkhout2019perfect} for use in sensing\cite{grigorenko1999phase, kabashin2009phase, zeng2015graphene, tsurimaki2018topological, sreekanth2018biosensing, kravets2013singular, wu2019layered} and optoelectronics\cite{ermolaev2021topological}.
We utilise the concept of cavity-free strong coupling\cite{munkhbat2019self, canales2020abundance, kaeek2021strong, thomas2021cavity}, where electromagnetic modes sustained by a material are strong enough to strongly couple to the material's own molecular resonance, to create phase singularities in a simple thin film of organic molecules. 
We show that the use of photochromic molecules allows for all-optical control of phase singularities.
Our results suggest a new application for strong light-matter coupling and a new, simplified, more versatile pathway to singular phase optics.
}

%%Main introduction...
%Along with amplitude and polarisation, phase is one of the fundamental properties of light\cite{hecht2002optics}.
Phase singularities occur when the amplitude of light reflected by an object is zero and its phase becomes undefined\cite{hecht2002optics}, leading to rapid changes in phase spectra\cite{grigorenko1999phase}.
Phase singularities could find application in sensing\cite{kabashin2009phase, zeng2015graphene, tsurimaki2018topological, sreekanth2018biosensing}, potentially enabling a sensitivity three orders of magnitude greater than commercial amplitude-based sensing technology\cite{kravets2013singular, wu2019layered},
and in flat optics, where it is otherwise difficult to induce strong optical phase variations in materials with deeply subwavelength thicknesses\cite{ermolaev2021topological}.
%These potential applications have motivated a search for nanostructures in which optical phase singularities can be observed.
Phase singularities have been observed using the Brewster angle\cite{hoenig1991direct}, surface plasmon resonances\cite{grigorenko1999phase, kabashin2009phase, zeng2015graphene, wu2019layered}, plasmonic lattices\cite{kravets2013singular, kravets2008extremely, malassis2014topological}, transition metal dichalcogenides\cite{ermolaev2021topological}, optical Tamm states\cite{tsurimaki2018topological} and Fabry-P\'{e}rot microcavities\cite{sreekanth2018biosensing, berkhout2019perfect}.
The realisation of phase singularities requires careful design of complicated structures using lithography\cite{kravets2013singular, kravets2008extremely, berkhout2019perfect}, self-assembly\cite{malassis2014topological} or multilayers\cite{tsurimaki2018topological, sreekanth2018biosensing, wu2019layered}.
In these structures, phase singularities can only be observed under a very specific set of conditions, such as one particular incident angle of light.
%For affordable applications in sensing and modulation, it would be desirable for phase singularities to be observable and the phase sensitivity easily tuneable in much simpler structures.

In this work, we show that strong light-matter coupling can be used to create phase singularities.
When an ensemble of molecular resonators is placed in a confined electromagnetic field, and if the molecular resonance and confined electromagnetic mode can be excited under the same conditions, the two modes can become coupled\cite{lidzey1998strong, garcia2021manipulating}.
When the coupling strength is sufficiently high the light and matter modes enter the strong coupling regime, forming hybrid states known as polaritons\cite{torma2014strong}.
While most strong coupling experiments rely on external structures (such as planar microcavities\cite{lidzey1998strong, schwartz2011reversible} or plasmonic nanostructures\cite{torma2014strong}) to generate confined electromagnetic fields, our results build on recent work which shows that such structures are not always needed\cite{munkhbat2019self, canales2020abundance, kaeek2021strong, thomas2021cavity}.
We use such a cavity-free design here to observe phase singularities associated with each newly created polariton state.
We use photochromic molecules to control the number of molecules, and hence light-matter coupling strength, by simple irradiation of light\cite{schwartz2011reversible}.
This allows one to tune into and detune from the phase singularities associated with each polariton branch, modifying the phase sensitivity of the films.
%We achieve this hitherto unrealised versatility in a remarkably simple organic thin film spin-coated on to a high-dielectric substrate.
Our results demonstrate what we believe to be the first all-optical control of phase singularities and the first application of strong light-matter coupling to the creation of phase singularities in a remarkably simple structure.

%Experiment

\begin{figure}[tb]
\includegraphics[scale=0.355]{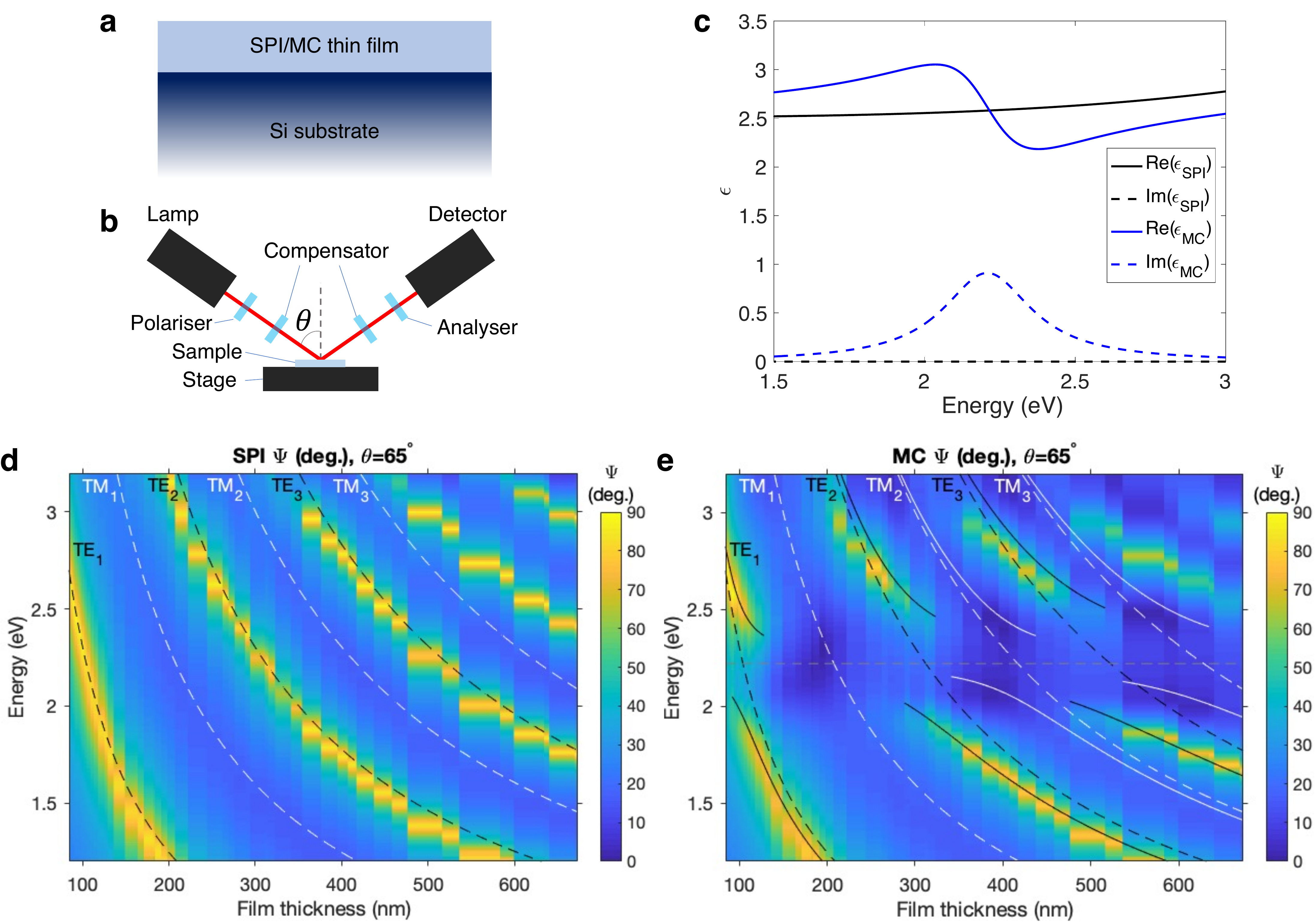}
\centering
\caption{\textbf{Cavity-free strong light-matter coupling.} (a) Sample design: SPI/MC film on silicon substrate.
(b) Ellipsometer schematic.
(c) Complex permittivities of SPI and MC, as derived from ellipsometry for a film of thickness 109 nm.
(d-e) Dispersion plots constructed using the ellipsometric parameter $\Psi$ for (d) SPI and (e) MC films over a range of thicknesses at fixed angle $\theta = 65^{\circ}$.
The dashed lines in (d-e) indicate the positions of the uncoupled TE (black) and TM (white) leaky modes.
The grey line in (e) at $E=2.22$ eV indicates the position of the MC molecular resonance.
The solid lines show the predicted positions of polariton branches using the $2N$ coupled oscillator model.
The coupled TE polariton branches (black) were fit with a coupling strength $g=225$ meV.
The TM$_2$ and TM$_3$ polariton branches (white) were fit with $g=185$ meV and $g=200$ meV, respectively.
We did not perform a coupled oscillator fit for the TM$_1$ mode since it shows no clear anticrossing.
}
\label{fig:design}
\end{figure}

We studied dielectric thin films of organic molecules deposited on silicon substrates by spin-coating (schematic in Figure \ref{fig:design}a; see Methods for fabrication details).
We used spiropyran (SPI), a molecule that is transparent to visible light and can be converted to merocyanine (MC) by exposure to ultraviolet light.
MC has a strong absorption peak at energy $E=E_{\text{MC}}=2.22$ eV (see Supplementary Information for SPI/MC molecular structures and MC transmission spectrum) and can be reconverted back into SPI after exposure to visible radiation\cite{berkovic2000spiropyrans}.
Reversible SPI/MC photoisomerisation has been exploited for all-optical control of strong coupling in Fabry-P\'{e}rot microcavities\cite{schwartz2011reversible} and lasing in plasmonic lattices\cite{taskinen2020all}.
The optical constants for SPI and MC are plotted in Figure \ref{fig:design}c.
They were determined using spectroscopic ellipsometry (see Methods).

We characterised our samples using spectroscopic ellipsometry (schematic in Figure \ref{fig:design}b; see also Methods).
Ellipsometry measures the complex reflection ratio $\rho$ in terms of the parameters $\Psi$ and $\Delta$\cite{tompkins2005handbook}:
\begin{align*}
    \rho = \frac{r_p}{r_s} = \tan(\Psi)e^{i\Delta}.
\end{align*}
$r_p$ and $r_s$ are the Fresnel amplitude reflection coefficients for p- and s-polarised light, respectively.
$\tan(\Psi)$ is the amplitude of $\rho$ and gives the ratio of the moduli of $r_p$ and $r_s$; $\Delta$ is the difference in the phase shifts undergone by p- and s-polarised light upon reflection.
%Ellipsometry simultaneously collects both amplitude and phase data.
Our ellipsometer's xenon light source emits low-intensity ultraviolet radiation, allowing us to monitor the conversion of SPI to MC.

In Figure \ref{fig:design}d we plot $\Psi$ spectra for SPI films with thicknesses in the range 84 nm $< L <$ 680 nm at a fixed incident angle $\theta=65^{\circ}$.
The large impedance mismatch between the Si substrate, SPI/MC thin films and air superstrate leads to the formation of leaky modes (also referred to as quasi-normal modes\cite{lalanne2018light}) in the film for both transverse electric (TE) and transverse magnetic (TM) polarised light.
For TE modes, $r_s < r_p$, giving maxima in $\Psi$; the positions of uncoupled TE leaky modes are indicated in Figure \ref{fig:design}d,e by dashed black lines.
For TM modes, $r_p < r_s$, giving minima in $\Psi$; the positions of uncoupled TM leaky modes are indicated in Figure \ref{fig:design}d,e by dashed white lines.
In Figure \ref{fig:design}e we plot $\Psi$ spectra for the films in Figure \ref{fig:design}d after photoconversion to MC.
Leaky modes can strongly couple to MC's molecular resonance.% at energy $E=E_{\text{MC}}=2.22$ eV.
There is clear anticrossing of the TE leaky modes around $E=E_{\text{MC}}$\cite{thomas2021cavity}, a signature of strong coupling between the TE leaky modes and MC resonance\cite{torma2014strong}.
%In fact, the coupling strength between the TE modes and MC resonance is high enough to put the system in the ultrastrong coupling regime\cite{thomas2021cavity}, where the coupling strength is comparable to the transition energy of the system.
TM modes are generally weaker than TE modes since $r_p$ is much lower than $r_s$ at the SPI/MC-Si interface\cite{hecht2002optics} at $\theta=65^{\circ}$.%, meaning that coupling between the TM modes and MC resonance is not high enough to fulfil the ultrastrong coupling regime.
A coupled oscillator fit (see Methods) suggests that the TM$_2$ and TM$_3$ modes have coupling strengths of $g_{\text{TM2}} = 185$ meV and $g_{\text{TM3}} = 200$ meV, meaning that they fulfil the strong coupling resolution criterion\cite{torma2014strong}:
\begin{align*}
    g_{\text{TM2,3}} > \frac{1}{4}\left(\gamma_{\text{MC}} + \gamma_{\text{TM2,3}}\right),
    %\label{eq:strong}
\end{align*}
\noindent where $\gamma_{\text{MC}} \approx 360$ meV is the MC resonance linewidth and $\gamma_{\text{TM2}} \approx 350$ meV and $\gamma_{\text{TM3}} \approx 240$ meV are the TM$_{2,3}$ leaky mode linewidths, respectively.
The TM$_1$ mode (for which $\gamma_{\text{TM1}} \approx 430$ meV) does not fulfil this criterion and there is no clear anti-crossing of the TM$_1$ mode in Figure \ref{fig:design}e.
%This suggests that the interaction between the TM$_1$ mode and the MC resonance falls within the weak coupling regime.

\begin{figure}[p]
\includegraphics[scale=0.35]{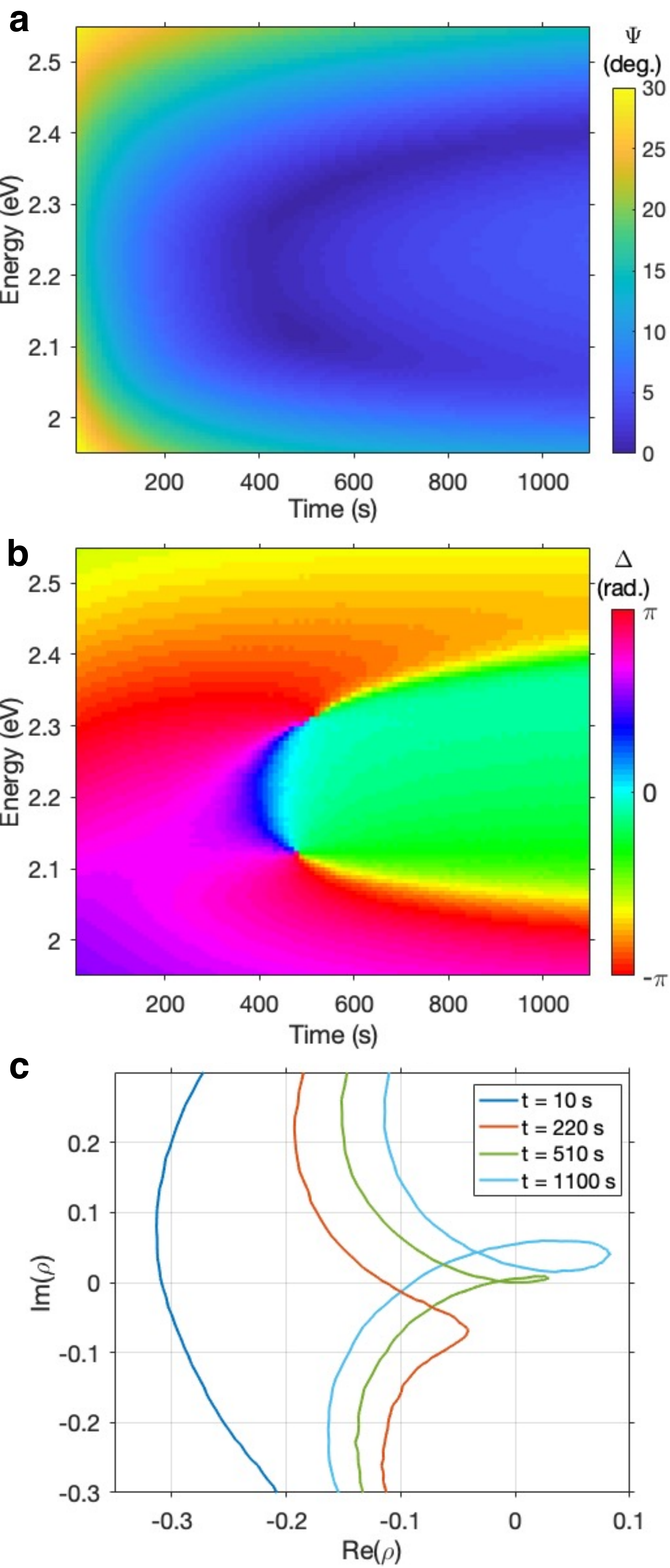}
\centering
\caption{\textbf{Creation of phase singularities in a SPI/MC thin film.}
The ellipsometric parameters (a) $\Psi$, (b) $\Delta$ and (c) $\rho$ plotted for a thin SPI film (thickness $L=407$ nm, measured at an incident angle $\theta=65^{\circ}$) while exposed to UV radiation.
As UV exposure time increases, SPI undergoes photoisomerisation to MC and the system transitions from the weak coupling regime to the strong coupling regime.
}
\label{fig:tm2}
\end{figure}

%Having established that all TE leaky modes and higher-order TM leaky modes in our dielectric film undergo strong coupling with the MC resonance, we study the evolution of the ellipsometric parameter $\rho$ for a single leaky as the film is converted from SPI to MC.
%Since $\rho \rightarrow \infty$ for TE phase singularities, we focus our analysis on the more straightforward case of TM phase singularities (for which $\rho=0$).
%We now show that phase singularities are created as a SPI/MC film enters the strong coupling regime.
In Figure \ref{fig:tm2} we show how the ellipsometric parameters $\Psi$, $\Delta$ and $\rho$ change as a thin SPI film is converted to MC.
For this thickness of film ($L=407$ nm) and incident angle of light ($\theta=65^{\circ}$) the energy of the TM$_2$ leaky mode matches that of the MC resonance.
Figure \ref{fig:tm2}a shows how $\Psi$ changes with increasing MC concentration.
After 400 seconds of UV exposure two distinct modes appear; as the MC concentration increases, the mode splitting increases.
This behaviour is consistent with the transition from the weak to strong coupling regime\cite{thomas2020new}.
The minimum value of $\Psi$ drops dramatically in the strong coupling regime, with one point on each branch going below $0.1^{\circ}$ (comparable with experimental error).
In Figure \ref{fig:tm2}b we plot $\Delta$, which shows the same phase behaviour between the two polariton branches that we have previously interpreted as a signature of strong coupling\cite{thomas2020new}.
Shortly after entering the strong coupling regime, we observe two phase singularities: the first, at $t=470$ s, on the lower polariton branch; the second, at $t=510$ s, on the upper polariton branch.
They correspond to positions on the upper and lower polariton branches in Figure \ref{fig:tm2}a at which $\Psi<0.1^{\circ}$.
The behaviour of these phase singularities is consistent with what has been previously reported\cite{grigorenko1999phase, kravets2013singular, ermolaev2021topological}: the phase cycles through $2\pi$ radians around each singularity, but in opposite directions to one another ($\Delta$ decreases when moving clockwise around the lower polariton phase singularity and increases when moving clockwise around the upper polariton phase singularity), indicating that the topological charge of the system has been preserved.

Figure \ref{fig:tm2}c shows $\rho$ at four times: $t=10$ s, when the film is almost entirely composed of SPI molecules; $t=220$ s, when some SPI has been converted into MC but the system remains in the weak coupling regime; $t=510$ s, coincident with the upper polariton phase singularity, when enough SPI has been converted to MC to place the system within the strong coupling regime; and $t=1100$ s, the final measurement when most SPI has been converted into MC.
The evolution of $\rho$ during the transition from weak to strong coupling resembles what we have observed in other strongly coupled systems\cite{thomas2020new}: the initial TM$_2$ leaky mode is represented by an arc; in the weak coupling regime this arc is perturbed by the presence of MC; in the strong coupling regime a new loop is observed.
As the MC concentration increases, the loop both increases in size and touches the point $\rho=0$ for $t=510$ s (green loop) and $t=470$ s (not plotted for clarity).
%At $t=510$ s, when $E=2.31$ eV, Re$(\rho) = (0.18 \pm 8.25) \times 10^{-2}$ and Im$(\rho) = (0.03 \pm 1.38) \times 10^{-2}$.
%This corresponds to a point on the lower side of the strong coupling loop.
%The phase singularity which we observe on the lower polariton branch at $t=470$ s occurs when the top of the strong coupling loop touches $\rho=0$: when $E=2.12$ eV, Re$(\rho) = (-0.03 \pm 3.15) \times 10^{-2}$ and Im$(\rho) = (-0.02 \pm 2.08) \times 10^{-2}$.
Provided the strong coupling loop passes over $\rho=0$ as the system evolves, we are guaranteed to observe two phase singularities.
%Similar analyses have been used to argue that phase singularities in other nanophotonic systems are topologically protected\cite{kravets2013singular, malassis2014topological, ermolaev2021topological}.

\begin{figure}[p]
\includegraphics[scale=0.6]{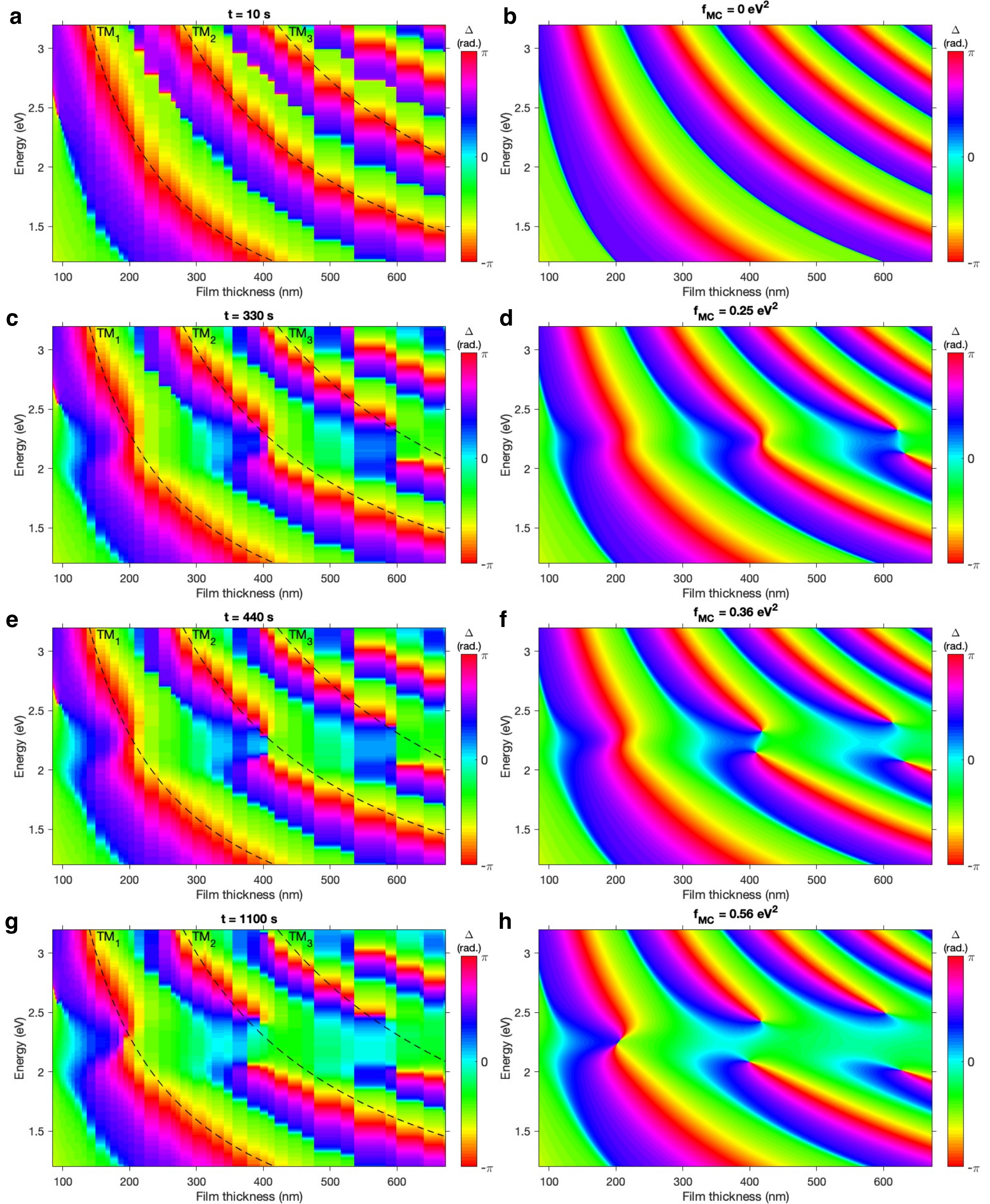}
\centering
\caption{\textbf{Creation of phase singularities for a range of film thicknesses.} (a,c,e,g) Experimental and (b,d,f,h) calculated (Fresnel approach) dispersion plots constructed using the ellipsometric parameter $\Delta$ for SPI/MC films over a range of thicknesses at fixed angle $\theta = 65^{\circ}$.
The MC concentrations were varied in experimental plots by UV exposure time ((a) $t=10$ s, (c) $t=330$ s, (e) $t=440$ s, (g) $t=1,100$ s) and in calculated plots by varying the Lorentz oscillator strength of MC ((b) $f_{\text{MC}} = 0$ eV$^2$, (d) $f = 0.25$ eV$^2$, (f) $f = 0.36$ eV$^2$ and (h) $f = 0.56$ eV$^2$).
The positions of the uncoupled TM modes are indicated by the black dashed lines in panels (a,c,e,g).
}
\label{fig:disp}
\end{figure}

%We now study the creation of phase singularities in SPI/MC films of a range of thicknesses.
In Figure \ref{fig:disp} we plot $\Delta$ for SPI/MC films with thicknesses 84 nm $< L <$ 680 nm.
Each film was exposed to UV radiation under the same conditions (with $\theta=65^{\circ}$).
We plot the phase response after (a) $t= 10$ s, (c) $t=330$ s, (e) $t=440$ s and (g) $t=1,100$ s of UV irradiation.
The dashed black lines indicate the positions of the uncoupled TM modes in the SPI films.
These results are well reproduced in Fresnel calculations (Figure \ref{fig:disp}(b,d,f,h)) by varying the Lorentz oscillator strength of the MC resonance.% while holding all other fit parameters constant.
We first observe the splitting of the TM$_3$ leaky mode (the best-confined of the plotted TM modes) in Figures \ref{fig:disp}c-d, followed by splitting of the TM$_2$ mode in Figures \ref{fig:disp}e-f.
As in Figure \ref{fig:tm2}b, topological charge is preserved as each pair of phase singularities is created.
We did not observe clear anti-crossing of the the TM$_1$ mode in Figure \ref{fig:design}e; likewise, here we do not unambiguously observe the creation of phase singularities associated with the TM$_1$ mode.
%We note that, as the number of MC molecules changes, the energy and film thickness of each phase singularity changes.
%A further degree of tuneability can be achieved by varying the angle of incidence; see \textbf{Supplementary Figure XXX}.
Phase singularities can also be created by coupling to TE modes under slightly different conditions; see Supplementary Section S2.

%Conclusion
In conclusion, we have shown that molecules in a thin film can undergo strong coupling to leaky modes in the same thin film of which they are a part, and that the transition from the weak to strong coupling regimes creates pairs of phase singularities.
The phase sensitivity of a thin film of a particular thickness can be varied by modifying the incident angle of light or the concentration of molecules in the film.
The use of photochromic molecules allowed us to demonstrate all-optical control of phase singularities.
Our results highlight a new application of strong coupling and a new, extraordinarily simple platform for singular phase optics.

\subsection*{Methods}

\subsection*{SPI/MC film fabrication}
Polymethyl methacrylate (PMMA, molar weight 450 000) was used as a host matrix for SPI (Sigma-Aldrich; see Supplementary Section 1 for SPI and MC molecular structures and MC transmission spectrum).
PMMA was dissolved in toluene. SPI was then dissolved in the PMMA-toluene solution with a weight ratio of 3:2 SPI to PMMA.
SPI/PMMA films were deposited on a silicon wafer by spin-coating three layers each with a spin speed of 2000 rpm.
This produced film thicknesses in the range 84--681 nm.

\subsection*{Optical constants of SPI and MC}
Optical constants of SPI and MC were determined using CompleteEASE\textregistered 6.51 (J.A. Woollam Co., Inc.).
The permittivity of SPI ($\epsilon_{\text{SPI}}$) was modelled as a Cauchy dielectric:
\begin{align*}
    \epsilon_{\text{SPI}}= \left( A+B\omega^2+C\omega^4 \right)^2
\end{align*}
where $A=1.584$, $B=-4.171 \times 10^{-34}$ rad$^{-2}$ s$^2$ and $C = 2.105 \times 10^{-64}$ rad$^{-4}$ s$^{4}$. The permittivity of MC ($\epsilon_{\text{MC}}$) was modelled with a Lorentz oscillator ($\epsilon_{L}$) and a pole in the ultraviolet ($\epsilon_{\text{UV}}$):  
\begin{align*}
    \epsilon_{\text{MC}} &= \epsilon_\infty + \epsilon_{\text{UV}} + \epsilon_{L} \\
    &= \epsilon_\infty + \frac{A_{\text{UV}}}{E_{\text{UV}}^2 - (\hbar\omega)^2} + \frac{f}{E_L^2 - (\hbar\omega)^2 - i\hbar\omega\gamma_L},
\end{align*}
where $\epsilon_\infty = 0.839$, $A_{\text{UV}} = 102.133$ eV$^2$, $E_{\text{UV}} = 7.970$ eV, $f=0.7226$ eV$^2$, $E_L = 2.215$ eV, $\gamma_L = 0.3587$ eV.

\subsubsection*{Ellipsometry}
Spectroscopic ellipsometry was carried out using a J. A. Woollam Co. M-2000XI which measures the ellipsometric parameters $\Psi$ and $\Delta$ in the wavelength range 210-1690 nm, with a wavelength step of 1.5 nm for 210-1000 nm and 3.5 nm for 1000-1690 nm. 
$\Psi$ gives the ratio of the field reflection coefficients for $p$- and $s$-polarised light (the moduli of $r_p$ and $r_s$, the complex Fresnel reflection coefficients for $p$- and $s$-polarised light  respectively) and $\Delta$ is the phase difference between the same coefficients such that $r_p/r_s = \tan(\Psi) e^{i\Delta}$.
Since ellipsometry measures the ratio of two quantities it is a very low-noise, sensitive technique.
This makes it possible to confidently identify points at which $\rho=0$ where we would expect to see phase singularities.

\subsubsection*{Coupled oscillator fit}
The lack of evidence for any mid-polariton bands in figure \ref{fig:design}e suggests that the polariton branches in our system are best modelled by a $2N$ coupling matrix\cite{richter2015maxwell}.
In this case the coupling between the MC resonance and each leaky mode can be described by the following matrix equation:
\begin{align*}
%\bigoplus_{j=1}^N
\left( \begin{array}{cc}
E_{\text{MC}}  & g_j \\
g_j & E_{j}
\end{array} \right) 
\left( \begin{array}{cc}
a_{\text{(L,U)}j} \\
b_{\text{(L,U)}j}
\end{array} \right)
 = E_{\text{(L,U)}j} \left( \begin{array}{cc}
a_{\text{(L,U)}j} \\
b_{\text{(L,U)}j}
\end{array} \right),
%\label{eq:2n}
\end{align*}
where %$N$ is the number of leaky modes couple to the TDBC resonance, 
$E_{\text{MC}}$ is the energy of the MC resonance, $g_j$ is the coupling strength between the MC resonance and the $j$th leaky mode, $E_{j}$ is the $j$th leaky mode energy, $E_{\text{(L,U)}j}$ are the energies of the lower and upper polariton branches associated with the $j$th leaky mode and $|a_{\text{(L,U)}j}|^2$ and $|b_{\text{(L,U)}j}|^2$ are the Hopfield coefficients describing the mixing of the MC resonance with the $j$th leaky mode.

\section*{Acknowledgements}
P.A.T. and W.L.B. acknowledge the support of the European Research Council through Project Photmat (ERC-2016-AdG-742222: www.photmat.eu).
K.S.M. and W. L. B acknowledge support from the Leverhulme Trust research grant ``Synthetic biological control of quantum optics''.
\newline Data in support of our findings are available at: https://ore.exeter.ac.uk/repository/handle/XXX

\section*{Author contributions}
P.A.T. and K.S.M. designed the experiment.
K.S.M. fabricated samples and P.A.T. performed measurements.
P.A.T. analysed experimental data.
P.A.T. and W.L.B. performed calculations.
P.A.T. wrote the manuscript with input from W.L.B. and K.S.M.

\section*{Competing interests}
The authors declare no competing interests.


\begin{thebibliography}{99}

\bibitem{lidzey1998strong}
	Lidzey, D. G. \textit{et al.}
	{Strong exciton--photon coupling in an organic semiconductor microcavity.}
	\textit{Nature}
	\textbf{395}, 53--55 (1998).
	DOI: 10.1038/25692

\bibitem{torma2014strong}
	T{\"o}rm{\"a}, P. \& Barnes, W. L.
	{Strong coupling between surface plasmon polaritons and emitters: a review.}
	\textit{Rep. Prog. Phys.}
	\textbf{78}, 013901 (2015).
	DOI: 10.1088/0034-4885/78/1/013901

\bibitem{garcia2021manipulating}
    Garc\'{i}a-Vidal, F. J., Ciuti, C. \& Ebbesen, T. W.
    {Manipulating matter by strong coupling to vacuum fields.}
    \textit{Science}
    \textbf{373}, eabd0336 (2021).
    DOI: 10.1126/science.abd0336

\bibitem{ramezani2017plasmon}
    Ramezani, M. \textit{et al.}
    {Plasmon-exciton-polariton lasing.}
    \textit{Optica}
    \textbf{4}, 31--37 (2021).
    DOI: 10.1364/OPTICA.4.000031

\bibitem{feist2017polaritonic}
	Feist, J., Galego, J. \& Garcia-Vidal, F. J.
	{Polaritonic chemistry with organic molecules.}
	\textit{ACS Photon.}
	\textbf{5}, 205--216 (2017).
	DOI: 10.1021/acsphotonics.7b00680

\bibitem{hoenig1991direct}
    Hoenig, D. \& Moebius, D.
    {Direct visualization of monolayers at the air-water interface by Brewster angle microscopy.}
    \textit{J. Phys. Chem.}
    \textbf{95}, 4590--4592 (1991).
    DOI: 10.1021/j100165a003

\bibitem{kravets2008extremely}
    Kravets, V. G., Schedin, F. \& Grigorenko, A. N.
    {Extremely Narrow Plasmon Resonances Based on Diffraction Coupling of Localized Plasmons in Arrays of Metallic Nanoparticles.}
    \textit{Phys. Rev. Lett.}
    \textbf{101}, 087403 (2008).
    DOI: 10.1103/PhysRevLett.101.087403

\bibitem{malassis2014topological}
    Malassis, L. \textit{et al.}
    {Topological Darkness in Self-Assembled Plasmonic Metamaterials.}
    \textit{Adv. Mater.}
    \textbf{26}, 324--330 (2014).
    DOI: 10.1002/adma.201303426

\bibitem{berkhout2019perfect}
    Berkhout, A. \& Koenderink, A. F.
    {Perfect Absorption and Phase Singularities in Plasmon Antenna Array Etalons.}
    \textit{ACS Photon.}
    \textbf{6}, 2917--2925 (2019).
    DOI: 10.1021/acsphotonics.9b01019

\bibitem{grigorenko1999phase}
    Grigorenko, A. N., Nikitin, P. I. \& Kabashin, A. V.
    {Phase jumps and interferometric surface plasmon resonance imaging.}
    \textit{Appl. Phys. Lett.}
    \textbf{75}, 3917 (1999).
    DOI: 10.1063/1.125493

\bibitem{kabashin2009phase}
    Kabashin, A. V., Patskovsky, S. \& Grigorenko, A. N.
    {Phase and amplitude sensitivities in surface plasmon resonance bio and chemical sensing.}
    \textit{Opt. Express}
    \textbf{17}, 21191--21204 (2009).
    DOI: 10.1364/OE.17.021191

\bibitem{zeng2015graphene}
    Zeng, S. \textit{et al.}
    {Graphene–Gold Metasurface Architectures for Ultrasensitive Plasmonic Biosensing.}
    \textit{Adv. Mater.}
    \textbf{27}, 6163--6169 (2015).
    DOI: 10.1002/adma.201501754

\bibitem{tsurimaki2018topological}
    Tsurimaki, Y. \textit{et al.}
    {Topological Engineering of Interfacial Optical Tamm States for Highly Sensitive Near-Singular-Phase Optical Detection.}
    \textit{ACS Photon.}
    \textbf{5}, 929--938 (2018).
    DOI: 10.1021/acsphotonics.7b01176

\bibitem{sreekanth2018biosensing}
    Sreekanth, K. V. \textit{et al.}
    {Biosensing with the singular phase of an ultrathin metal-dielectric nanophotonic cavity.}
    \textit{Nat. Commun.}
    \textbf{9}, 369 (2018).
    DOI: 10.1038/s41467-018-02860-6

\bibitem{kravets2013singular}
    Kravets, V. G. \textit{et al.}
    {Singular phase nano-optics in plasmonic metamaterials for label-free single-molecule detection.}
    \textit{Nat. Mater.}
    \textbf{12}, 304--309 (2013).
    DOI: 10.1038/nmat3537

\bibitem{wu2019layered}
    Wu, F. \textit{et al.}
    {Layered material platform for surface plasmon resonance biosensing.}
    \textit{Sci. Rep.}
    \textbf{9}, 20286 (2019).
    DOI: 10.1038/s41598-019-56105-7
    
\bibitem{ermolaev2021topological}
    Ermolaev, G. \textit{et al.}
    {Topological phase singularities in atomically thin high-refractive-index materials.}
    Preprint at https://arxiv.org/abs/2106.12390 (2021).

\bibitem{munkhbat2019self}
    Munkhbat, B. \textit{et al.}
    {Self-hybridized exciton-polaritons in multilayers of transition metal dichalcogenides for efficient light absorption.}
    \textit{ACS Photon.}
    \textbf{6}, 139--147 (2019).
    DOI: 10.1021/acsphotonics.8b01194

\bibitem{canales2020abundance}
    Canales, A., Baranov, D. G., Antosiewicz, T. J. \& Shegai, T.
    {Abundance of cavity-free polaritonic states in resonant materials and nanostructures.}
    \textit{J. Chem. Phys.}
    \textbf{154}, 024701 (2021).
    DOI: 10.1063/5.0033352

\bibitem{kaeek2021strong}
    Kaeek, M., Damari, R., Roth, M., Fleischer, S. \& Schwartz, T.
    {Strong Coupling in a Self-Coupled Terahertz Photonic Crystal.}
    \textit{ACS Photon.}
    \textbf{8}, 1881--1888 (2021).
    DOI: 10.1021/acsphotonics.1c00309

\bibitem{thomas2021cavity}
    Thomas, P. A., Menghrajani, K. S. \& Barnes, W. L.
    {Cavity-Free Ultrastrong Light-Matter Coupling.}
    \textit{J. Phys. Chem. Lett.}
    \textbf{12}, 6914--6918 (2021).
    DOI: 10.1021/acs.jpclett.1c01695

\bibitem{hecht2002optics}
    Hecht, E.
    \textit{Optics} (5th edition, Addison Wesley, 2002.)


    


\bibitem{schwartz2011reversible}
	Schwartz, T., Hutchison, J. A., Genet, C. \& Ebbesen, T. W.
	{Reversible switching of ultrastrong light-molecule coupling.}
	\textit{Phys. Rev. Lett.}
	\textbf{106}, 196405 (2011).
	DOI: 10.1103/PhysRevLett.106.196405



\bibitem{berkovic2000spiropyrans}
	Berkovic, G., Krongauz, V. \& Weiss, V.
	{Spiropyrans and spirooxazines for memories and switches.}
	\textit{Chem. Rev.}
	\textbf{100}, 1741--1754 (2000).
	DOI: 10.1021/cr9800715

\bibitem{taskinen2020all}
    Taskinen, J. M. \textit{et al.}
    {All-Optical Emission Control and Lasing in Plasmonic Lattices.}
    \textit{ACS Photonics}
    \textbf{7}, 2850--2858 (2020).
    DOI: 10.1021/acsphotonics.0c01099

\bibitem{tompkins2005handbook}
	Tompkins, H. \& Irene, E. A.
	\textit{Handbook of Ellipsometry.}
	(William Andrew Publishing, 2005.)

\bibitem{lalanne2018light}
    Lalanne, P., Yan, W., Vynck, K., Sauvan, C. \& Hugonin, J.-P.
    {Light Interaction with Photonic and Plasmonic Resonances.}
    \textit{Laser Photonics Rev.}
    \textbf{12}, 1700113 (2018).
    DOI: 10.1002/lpor.201700113



\bibitem{richter2015maxwell}
    Richter, S. \textit{et al.}
    {Maxwell consideration of polaritonic quasi-particle Hamiltonians in multi-level systems.}
    \textit{Appl. Phys. Lett.}
    \textit{107}, 231104 (2015).
    DOI: 10.1063/1.4937462

\bibitem{menghrajani2020strong}
    Menghrajani, K. S. \& Barnes, W. L.
    {Strong Coupling beyond the Light-Line.}
    \textit{ACS Photon.}
    \textit{7}, 2448–2459 (2020).
    DOI: 10.1021/acsphotonics.0c00552

\bibitem{thomas2020new}
    Thomas, P. A., Tan, W. J., Fernandez, H. A. \& Barnes, W. L.
    {A New Signature for Strong Light--Matter Coupling Using Spectroscopic Ellipsometry.}
    \textit{Nano Lett.}
    \textit{20}, 6412--6419 (2020).
    DOI: 10.1021/acs.nanolett.0c01963
    







\end{thebibliography}
\end{document}

\maketitle

\tableofcontents

\newpage

\addcontentsline{toc}{section}{S1. SPI/MC chemical structure and MC transmittance spectrum}
\section*{S1. SPI/MC chemical structure and MC transmittance spectrum}
\renewcommand{\thefigure}{S1}
\begin{figure}[!h]
\includegraphics[scale=0.60]{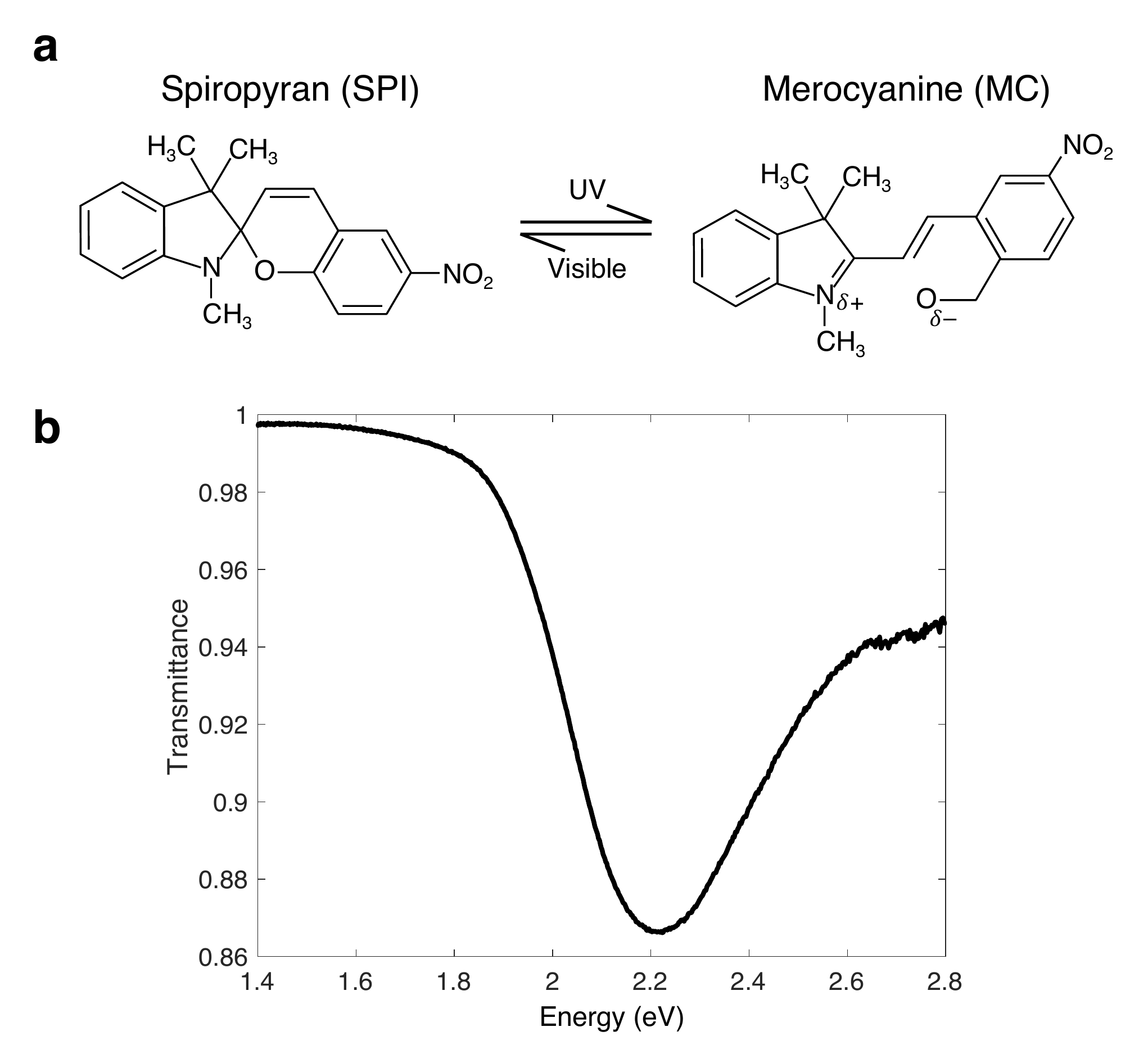}
\centering
\caption{(a) Spiropyran and merocyanine chemical structures. (b) Transmittance through a merocyanine (MC) film (thickness 150 nm) spin-coated on a glass substrate, normalised against transmission for an uncoated substrate (see Methods for fabrication).
}
\end{figure}

\newpage
\addcontentsline{toc}{section}{S2. Emergence of phase singularities for strongly coupled TE modes}
\section*{S2. Emergence of phase singularities for strongly coupled TE modes}

In the main manuscript we observed the creation of phase singularities when TM leaky modes strongly coupled to the MC molecular resonance.
We did not observe the creation of new phase singularities arising from the strong coupling of TE leaky modes and the MC molecular resonance.

Here we seek to answer two questions:
\begin{enumerate}
    \item Why did we observe the creation of phase singularities associated with strongly coupled TM modes but not with TE modes? (Even though the coupling between TE modes and the MC resonance is stronger than between TM modes and the MC resonance.)
    \item Is it possible to create phase singularities by coupling TE leaky modes to the MC resonance?
\end{enumerate}

To answer the first question, in Figure S2 we plot ellipsometric parameters (calculated using the transfer matrix method) for SPI films over a wide energy range while varying the angle of incidence $\theta$.
At $\theta=15^{\circ}$ (Fig. S2a-b) we observe phase singularities for each TE leaky mode.
At such a shallow incident angle the phase singularities in each pair occur at very similar energies; their energies closely match the energy ($E=4.45$ eV) at which our Cauchy model for SPI predicts that the SPI film and Si substrate permittivities match ($\epsilon_{\text{SPI}} = \epsilon_{\text{Si}}$).
Increasing the incident angle to $\theta=40^{\circ}$ (Fig. S2c-d) increases the separation between each pair of phase singularities: the higher energy phase singularities are shifted to even higher energies and the lower energy phase singularities are shifted to even lower energies.
Increasing $\theta$ from $40^{\circ}$ to $60^{\circ}$ (Fig. S2e-f) shifts the higher energy phase singularities outside our plotted range while the lower energy phase singularities have been shifted to 2.8 eV.
Increasing $\theta$ to $65^{\circ}$ shifts the lower phase singularity outside our plotted range.
These results are consistent with our experimental results in Main Figure 3a (for which $\theta=65^{\circ}$) to within a few degrees.
The slight mismatch between experiment and calculation in predicting the phase response is likely due to our choice of model of the permittivity of SPI.
Our simple Cauchy model (see Methods) accurately describes the behaviour of SPI well in the spectral region 1.2 eV $<E<$ 3.2 eV (apart from the precise conditions for the phase singularity), but is inadequate in describing its UV response and hence predicting where $\epsilon_{\text{SPI}} = \epsilon_{\text{Si}}$.

To summarise our answer to the first question, we did not observe the creation of new phase singularities arising from the strong coupling of TE leaky modes and the MC molecular resonance because at $\theta=65^{\circ}$ pairs of TE phase singularities already exist in SPI films but cannot be fully observed in the plotted experimental range in Main Figure 3.

\renewcommand{\thefigure}{S2}
\begin{figure}[!p]
\includegraphics[scale=0.55]{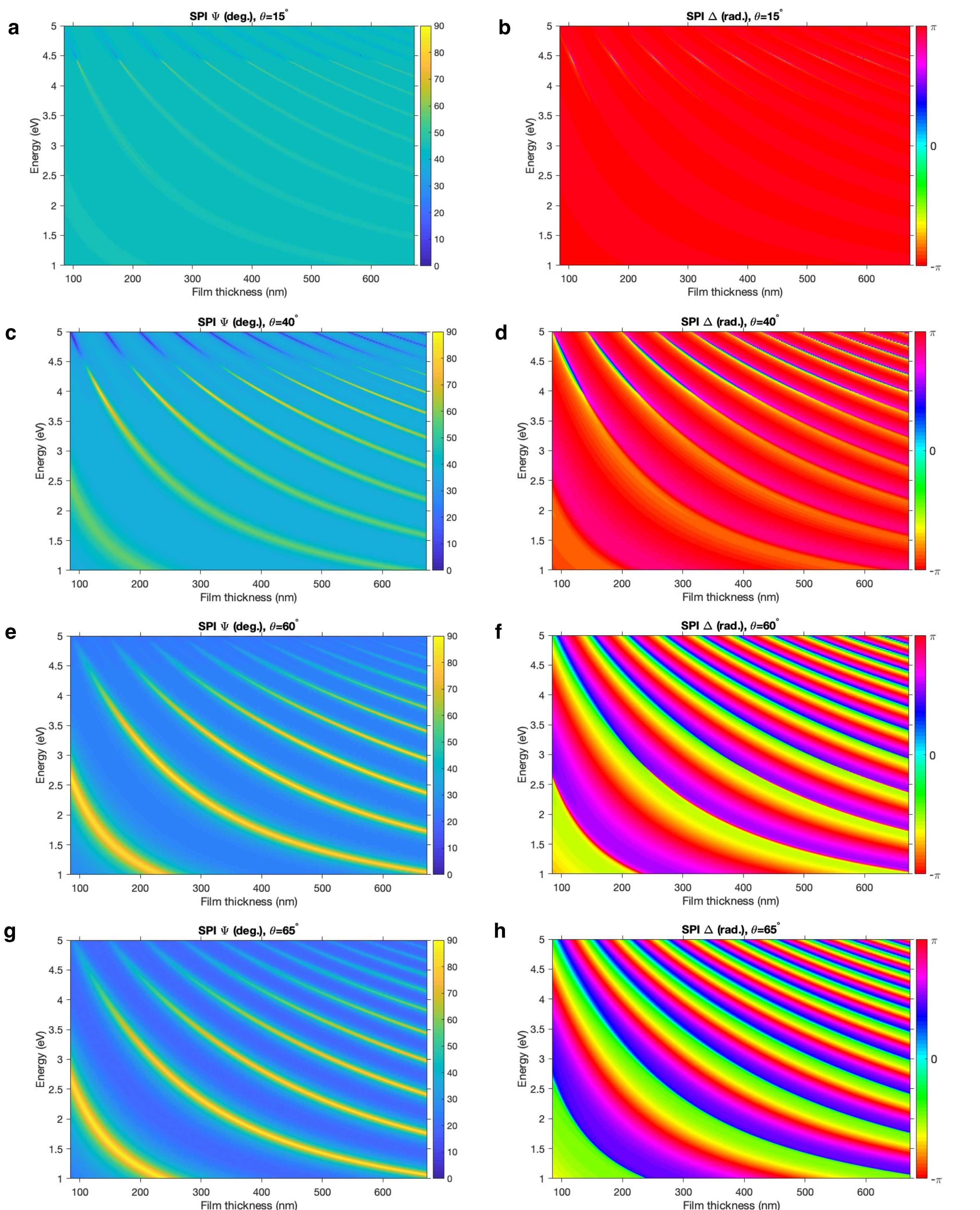}
\centering
\caption{
Calculated ellipsometric parameters $\Psi$ and $\Delta$ for SPI films over a range of thicknesses for incident angles of (a,b) $15^{\circ}$, (c,d) $40^{\circ}$, (e,f) $60^{\circ}$ and (g,h) $65^{\circ}$.
}
\end{figure}

\newpage
To answer the second question, we first note that we conducted our experiments at a high incident angle of $\theta=65^{\circ}$ to generate the confined electromagnetic fields necessary for the creation of TM phase singularities.
At such a high incident angle, pairs of TE phase singularities have already been created but cannot be fully observed in the plotted experimental range in Main Figure 3.
TE leaky modes generate stronger confined electromagnetic fields than TM leaky modes and electromagnetic field confinement decreases with lower incident angle\cite{hecht2002optics, thomas2021cavity}.
We therefore expect to see the creation of phase singularities associated with strong coupling between TE modes and the MC resonance at a lower incident angle.
This is confirmed in Figure S3, where we have calculated the ellipsometric parameters for MC films at $\theta=40^{\circ}$ and varied the MC resonance Lorentz oscillator strength.
There are many similarities to the phase response of TM modes as they undergo strong coupling (Manuscript Figure 3).
As the Lorentz oscillator strength of the MC resonance is increased, pairs of phase singularities appear as each TE mode couples to the MC resonance.
Higher order TE leaky modes (which match the MC resonance for thicker films) are better confined than lower order leaky modes.
Therefore, as we gradually increase oscillator strength we first observe the anticrossing indicative of strong coupling in $\Psi$ (and the corresponding phase singularity creation in $\Delta$) for higher-order TE modes.
However, unlike the TM case in Main Figure 3, here we observe the creation of two pairs of phase singularities as each TE mode strongly couples with the MC resonance.
We first observe the creation of a phase singularity pair as each TE leaky mode is first perturbed by the MC resonance in the weak coupling regime.
Then, as anticrossing becomes clear in $\Psi$, a second pair of phase singularities appears.
These second pairs of phase singularities, approximately corresponding to the ends of each polariton branch at the strong coupling-induced photonic stop-band, behave in the same manner as the phase singularities observed in Main Figure 3.

To summarise our answer to the second question, it is possible to create phase singularities by coupling TE leaky modes to the MC resonance when the angle of incidence is decreased. When TE modes strongly couple to the MC resonance it is possible to observe the creation of two pairs of phase singularities per leaky mode (instead of just one pair per leaky mode, as is the case for TM modes).

\renewcommand{\thefigure}{S3}
\begin{figure}[!p]
\includegraphics[scale=0.55]{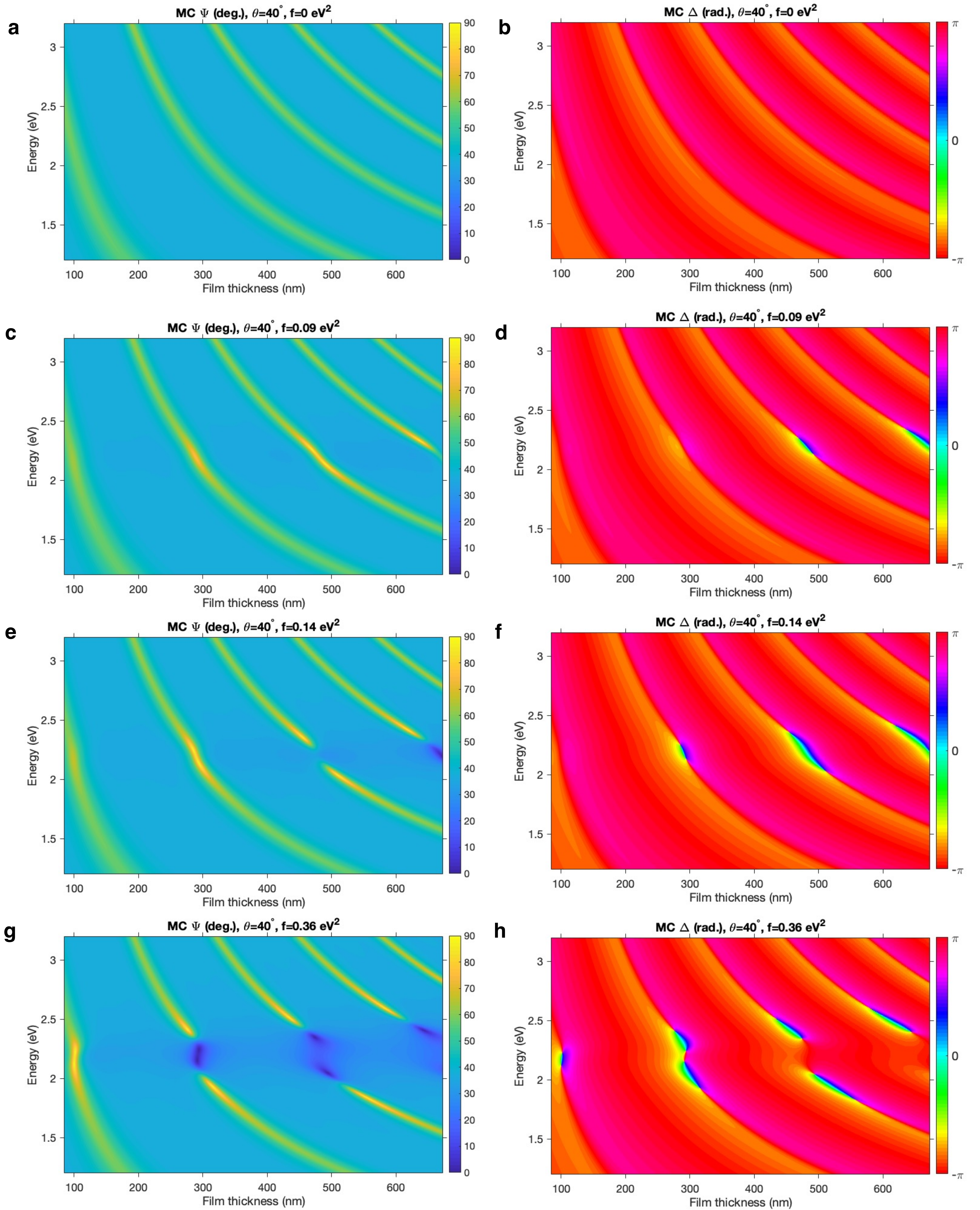}
\centering
\caption{
Calculated ellipsometric parameters $\Psi$ and $\Delta$ for MC films over a range of thicknesses at an incident angles of $40^{\circ}$ with the MC Lorentz oscillator strength set to (a,b) 0 eV$^2$, (c,d) 0.09 eV$^2$, (e,f) 0.14 eV$^2$ and (g,h) 0.36 eV$^2$.
}
\end{figure}